\documentstyle[12pt]{article}

\renewcommand{\baselinestretch}{1}
\def\address{\m@th\@ifnextchar[\@address{\@address[]}}
 
\def\@address[#1]#2{
\expandafter\def\expandafter\@addressname\expandafter
{\@addressname{
  \adr{#1}\ \parbox[t]{5in}{
     \ignorespaces #2}\par }}}
\def\@addressname{}
\def\adr#1{{\normalsize\unskip$^{#1}$}}
\def\@maketitle{%
\def\and{{\rm and}}
  \newpage
  \null
  {\centering
  \let \footnote \thanks
    {\Large\bf   \@title \par}%
    \vskip 1.5em%
      \lineskip .5em%
    {\bf\normalsize   \@author\par}
     \vspace{1em} 
    {\small \@addressname}
    
  }%
  \par
  \vskip 1.5em}

\title{ \vspace*{-8mm}\Large \bf Novel algebraic structures from the 
polysymplectic form in field theory\thanks{to appear in: 
{GROUP21, \it Physical Applications and Mathematical Aspects of
Geometry, Groups, and Algebras,
}
Vol. 2, eds. H.-D. Doebner, W. Scherer, C. Schulte, 
(World Scientific, Singapore, 1997), p. 894
} }
\author{ \vspace*{-7mm} \\
\rm Igor V. Kanatchikov\thanks{E-mail address: kai@fuw.edu.pl}\\
\small \it Center of Theoretical Physics,
\small  \it Polish Academy of Sciences \\ \vspace*{-10pt} 
\small \it Al. Lotnik\'ow 32/46, 
\small  \it  PL-02-668 Warszawa, Poland   
 }
\date{}

\begin{document}

\maketitle

\vspace*{-72mm} 
\begin{flushright}
hep-th/9612255
 \end{flushright}
\vspace*{53mm}


\newcommand{\beq}{\begin{equation}}
\newcommand{\eeq}{\end{equation}}
\newcommand{\beqa}{\begin{eqnarray}}
\newcommand{\eeqa}{\end{eqnarray}}
\newcommand{\nn}{\nonumber}

\newcommand{\half}{\frac{1}{2}}

\newcommand{\xt}{\tilde{X}}

\newcommand{\uind}[2]{^{#1_1 \, ... \, #1_{#2}} }
\newcommand{\lind}[2]{_{#1_1 \, ... \, #1_{#2}} }
\newcommand{\com}[2]{[#1,#2]_{-}} 
\newcommand{\acom}[2]{[#1,#2]_{+}} 
\newcommand{\compm}[2]{[#1,#2]_{\pm}}

\newcommand{\lie}[1]{\pounds_{#1}}
\newcommand{\co}{\circ}
\newcommand{\sgn}[1]{(-)^{#1}}
\newcommand{\lbr}[2]{ [ \hspace*{-1.5pt} [ #1 , #2 ] \hspace*{-1.5pt} ] }
\newcommand{\lbrpm}[2]{ [ \hspace*{-1.5pt} [ #1 , #2 ] \hspace*{-1.5pt}
 ]_{\pm} }
\newcommand{\lbrp}[2]{ [ \hspace*{-1.5pt} [ #1 , #2 ] \hspace*{-1.5pt} ]_+ }
\newcommand{\lbrm}[2]{ [ \hspace*{-1.5pt} [ #1 , #2 ] \hspace*{-1.5pt} ]_- }
\newcommand{\pbr}[2]{ \{ \hspace*{-2.2pt} [ #1 , #2 ] \hspace*{-2.55pt} \} }
\newcommand{\we}{\wedge}
\newcommand{\dv}{d^V}
\newcommand{\nbrpq}[2]{\nbr{\xxi{#1}{1}}{\xxi{#2}{2}}}
\newcommand{\lieni}[2]{$\pounds$${}_{\stackrel{#1}{X}_{#2}}$  }

\newcommand{\rbox}[2]{\raisebox{#1}{#2}}
\newcommand{\xx}[1]{\raisebox{1pt}{$\stackrel{#1}{X}$}}
\newcommand{\xxi}[2]{\raisebox{1pt}{$\stackrel{#1}{X}$$_{#2}$}}
\newcommand{\ff}[1]{\raisebox{1pt}{$\stackrel{#1}{F}$}}
\newcommand{\dd}[1]{\raisebox{1pt}{$\stackrel{#1}{D}$}}
\newcommand{\nbr}[2]{{\bf[}#1 , #2{\bf ]}}
\newcommand{\der}{\partial}
\newcommand{\oo}{$\Omega$}
\newcommand{\Om}{\Omega}
\newcommand{\om}{\omega}
\newcommand{\eps}{\epsilon}
\newcommand{\si}{\sigma}
\newcommand{\Lm}{\bigwedge^*}

\newcommand{\inn}{\hspace*{2pt}\raisebox{-1pt}{\rule{6pt}{.3pt}\hspace*
{0pt}\rule{.3pt}{8pt}\hspace*{3pt}}}
\newcommand{\sro}{Schr\"{o}dinger\ }
\newcommand{\bm}{\boldmath}
\newcommand{\vol}{\omega}

\newcommand{\bd}{\mbox{\bm $d$}}
\newcommand{\bder}{\mbox{\bm $\der$}}
\newcommand{\bI}{\mbox{\bm $I$}}


\vspace*{-2mm}

\begin{abstract} 
The polysymplectic $(n+1)$-form is introduced as an analogue 
of the symplectic form for the De Donder-Weyl polymomentum 
Hamiltonian formulation of field theory.   
The corresponding Poisson brackets on differential forms
are constructed. 
The analogues of the Poisson algebra are shown to be generalized 
(non-commutative and higher-order) Gerstenhaber algebras 
defined in the text. 
\end{abstract}
\noindent

In classical field theory there exists an interesting formulation of 
 field equations which resembles Hamilton's equations in mechanics, 
but does not require a singling out of the time variable as in  the 
standard Hamiltonian formalism. 
This is the so-called DeDonder-Weyl (DW) 
formulation (see e.g. \cite{ref1}). 
Known from 
the calculus of variations since the thirties it still has no wide 
applications because generalizations 
 of the powerful structures of the 
canonical formalism in mechanics to this approach are 
rather poorly understood.  
This short report concisely presents recent results of the author 
in this field. 
Proofs and details are to be given elsewhere (see also \cite{ikanat}).


Let $X$ is the space-time manifold and 
fields are viewed as sections of the {\em covariant 
configuration bundle} 
$\pi_{XY}:Y\rightarrow X$,
with coordinates
$(y^a,x^i)$, $i=1,...,n$, $a=1,...,m$, where $y^a$ denote the 
field variables. A field theory is given by 
the Lagrangian  function $L=L(y^a, y^a_i,x^i)$ on the first jet bundle  
of $Y$, $J^1Y$, with coordinates $(y^a, y^a_i,x^i)$.
Classical field configurations are critical sections 
$\sigma:X\rightarrow Y$ for which $\delta \int_X L\omega =0$, where 
$\omega:=dx^1\wedge ... \wedge dx^n $ is 
the volume $n$-form on $X$. 
These critical sections are known to be spanned by vector fields 
$\xi\in TJ^1Y$ 
which annihilate the exterior differential of the Lagrangian 
{\em Poincar\'e-Cartan form} on $J^1Y$ given by (see e.g. \cite{pcform}) 
$\Omega_L=dy^a\wedge d (\frac{\der L}{\der y^a_i})\wedge \der_i\inn \omega
-d(\frac{\der L}{\der y^a_i} y^a_i - L) \wedge \omega, 
$
that is 
$(j_1 \sigma)^*(\xi\inn\Omega_L)=0.
$ 
This condition is known to be 
equivalent to the familiar Euler-Lagrange equations.

To obtain  an analogue of the Hamiltonian formulation 
let us introduce the {\em polymomenta} $p^i_a:=\frac{\der L}{\der y^a_i}$ 
and the {\em DW Hamiltonian function} 
$H(y^a, p^i_a, x^i):= p_a^i y^a_i (y^a, p^i_a, x^i) - L$. 
Here one  assumes that $det||\der^2 L/ \der y^a_i \der y^b_j || \neq 0$, 
    \renewcommand{\baselinestretch}{1} 
which in general is different from the  regularity condition in 
standard instantaneous Hamiltonian formalism. In terms of  the 
variables above  
the (differential of the) Poincar\'e-Cartan form takes the form 
\beq
\Omega_{PC}=dy^a\wedge dp^i_a\wedge\omega_i-dH\wedge\omega,  
\eeq
where $\omega_i:=\der_i\inn \omega$. 
Again,  isotropic subspaces of $\Omega_{PC}$ give rise to the 
equations of motion, this time in  first order (i.e.  Hamiltonian)  
formulation (see e.g. [3]): 
\beq
\der_i p^i_a=-\der H / \der y^a ,
\quad \der_i y^a = \der H / \der  p^i_a .
\eeq
This is the  DeDonder-Weyl Hamiltonian form of  field equations.  
It is entirely space-time symmetric -- i.e. no distinction  
is made between the time and space dimensions; 
and its arena is 
the finite dimensional 
(extended) 
{\em polymomentum  phase space} of variables 
$z^M=(y^a, p^i_a,x^i)$. 
These features make this formulation 
particularly interesting. 
 
Let us consider  how the structures of the 
Hamiltonian formalism in mechanics can be generalized 
to the DW formulation of field theory. 
First 
we construct an
 analogue of the symplectic form. 
Let $\bigwedge^p_q(Y)$ 
denotes the 
bundle 
of $(p-q)$-horizontal $p$-forms on $Y$, whose fiber over $y\in Y$ 
is the space 
$\bigwedge^p_{q (y)}$ of forms
which  are annihilated by $(q+1)$ 
arbitrary vertical vectors of $T_yY$: 
$\theta\in \bigwedge^p_{q(y)} \subset 
\bigwedge^p_{(y)}$ iff $v_1\inn ... \inn ... v_{q+1}\inn \theta=0$ for all 
$v_1, ... ,  v_{q+1}\in V_yTY$.
Consider the 
fibered manifold 
$\pi_{Y{\cal Z}}: {\cal Z} \rightarrow Y$ 
whose fiber over $y\in Y$ is the  
coset ${\cal Z}_{(y)}:=\bigwedge^n_{1(y)}/ \bigwedge^n_{0(y)}$. 
The elements of ${\cal Z}$ are 
\beq
\Theta = - p_a^i dy^a \wedge \omega_i , 
\eeq
where $p_a^i$ are the fiber coordinates for ${\cal Z}$, and 
$\Theta$ is regarded as a coset rather than an $n$-form on any manifold. 
The composition 
$\pi_{XY}\co \pi_{Y{\cal Z}}:=\pi_{X{\cal Z} }: {\cal Z}\rightarrow X $ 
yields the fibered manifold corresponding to the extended polymomentum phase 
space.
Define 
the vertical exterior 
differential $\dv$: 
$\bigwedge^p_q({\cal Z})\rightarrow\bigwedge^{p+1}_{q+1}({\cal Z})$ 
so that 
for all $\theta \in \bigwedge^p_q({\cal Z})$, $\dv \theta := 
[d\theta \; \; mod(\bigwedge^{p+1}_q({\cal Z})]$, where 
$d$: $\bigwedge^p_q({\cal Z})\rightarrow\bigwedge^{p+1}_{q+1}({\cal Z})$ 
is the usual 
exterior differential on $\bigwedge^*({\cal Z})$, 
and $[\; \;]$ denotes an equivalence class.
Now,
our basic 
object 
on the polymomentum phase 
space ${\cal Z}$, 
the {\em polysymplectic form},  
is defined  as follows
\beq
\Omega := \dv \Theta = - dp^i_a \wedge dy^a \wedge \omega_i , 
\eeq
which is again understood as 
a representative of the equivalence class. In fact, 
for all $z\in {\cal Z}$,  
$\Omega_{(z)}=[\Omega_{PC}{}_{(z)} \; \; mod \bigwedge^{n+1}_{1}{}_{(z)}]$ 
(cf. eq. (1)). 

The polysymplectic form generalizes the symplectic form of mechanics and 
reduces to the latter as $n=1$. 
Recall that the symplectic form determines  
the map between  dynamical variables (= functions on the phase space)  
and Hamiltonian vector fields which gives rise to the definition of the 
Poisson bracket of 
dynamical variables. Here, the polysymplectic form can be used for 
mapping 
of 
horizontal forms $\ff{p}\in \bigwedge^p_0({\cal Z})$, 
$\ff{p}=\frac{1}{p!}F\lind{i}{p}(z) dx\uind{i}{p}$, 
which will play the role of dynamical variables, to {\em vertical} 
multivector fields of degree $(n-p)$, $\xx{n-p}$: 
\beq
\xx{n-p}{}_F\inn \Omega = \dv \ff{p} ,
\eeq   
where the multivector field of degree $q$ on ${\cal Z}$: 
$\xx{q}\in \bigwedge^q(T{\cal Z})$ is said to be {\em vertical} if it 
annihilates any horizontal $q$-form:  
$\xx{q}\inn \ff{q}=0$ for all $\ff{q}\in\bigwedge^p_0({\cal Z})$. 
The space of vertical $q$-multivectors will be denoted 
$V\! \bigwedge^q(T{\cal Z})$. 

Using  (5) one can define the Poisson bracket on 
horizontal  forms: 
\beq
\pbr{\ff{p}{}_1}{\ff{q}{}_2}:= (-)^{n-p}X_{F_1}\inn X_{F_2} \inn \Omega.
\eeq
This bracket is graded antisymmetric 
and fulfills the graded Jacobi identity
\beq
\pbr{\ff{p}_1}{\ff{q}_2} = -(-1)^{g_1 g_2}
\pbr{\ff{q}_2}{\ff{p}_1}, 
\eeq
\beq
\mbox{$(-1)^{g_1 g_3} \pbr{\ff{p}}{\pbr{\ff{q}}{\ff{r}}}$} 
+
\mbox{$(-1)^{g_1 g_2} \pbr{\ff{q}}{\pbr{\ff{r}}{\ff{p}}}$} 
+
\mbox{$(-1)^{g_2 g_3} \pbr{\ff{r}}{\pbr{\ff{p}}{\ff{q}}}= 0,$}  
\eeq
where $g_1 = n-p-1$, $g_2 = n-q-1$ and $g_3 = n-r-1$.  

In addition to this graded Lie algebra structure we have here a higher-order 
graded analogue of the Poisson property of the bracket. This is seen from the 
fact that the multivector field of degree $(n-p)$ one associates with a 
form of 
degree $p$ is a graded differential operator 
(g.d.o., for short) 
of order  $(n-p)$ on the 
exterior algebra.  
The corresponding 
{\em higher-order} graded Leibniz rule can be written  in terms of the 
$r$-linear 
maps $\Phi^r_D$ associated with a 
g.d.o. $D$ on the 
exterior (in fact, any graded commutative) algebra $\bigwedge^*$, 
$\Phi{}^r_D: \bigotimes{}^r \Lm \rightarrow \Lm$, 
which were introduced by Koszul \cite{koszul}. 
By definition, 
\beq
\Phi{}^r_D(F_1, ..., F_r) := 
m \co (D\otimes {\mbox{\bf 1}}) 
\lambda{}^r (F_1\otimes ... \otimes F_r),
\eeq
for all $F_1, ..., F_r$ in $\Lm$, where $m$ is the multiplication map in 
$\Lm$,  $m(F_1\otimes F_2):=F_1 \we F_2$; $\lambda{}^r$ is a linear map 
$\bigotimes{}^r \Lm \rightarrow \Lm \otimes \Lm$: 
$\lambda{}^r (F_1\otimes ... \otimes F_r) :=
\lambda(F_1) \we ... \we \lambda(F_r) $; 
and the map 
$\lambda: \Lm \rightarrow \Lm \otimes \Lm $ is given by 
$\lambda(F) := F\otimes {\mbox{\bf 1}} - {\mbox{\bf 1}} \otimes F$.
The graded differential operator $D$ is said to be of $r$-th 
order iff $\Phi{}^{r+1}_D = 0$ identically. 

The higher-order Leibniz rule for the (left) bracket with a $p$-form
is written now as follows:
\beq
{\mbox{\Large $\Phi$}}^{n-p+1}_{\mbox{\small $\pbr{\ff{p}}{\,.\,} $} }
(F_1, ... , F_{n-p+1})=0.
\eeq
If $p=(n-1)$ this reproduces the usual graded Leibniz rule. 
The simplest non-trivial
 generalization corresponds to $p=(n-2)$. In this case we obtain the following 
second-order graded Leibniz rule 
\beqa
\pbr{\ff{n-2}}{\ff{q}\we\ff{r}\we\ff{s}}
&=&\pbr{\ff{n-2}}{\ff{q}\we\ff{r}}\we\ff{s} 
+(-)^{q(r+s)}\pbr{\ff{n-2}}{\ff{r}\we\ff{s}}\we\ff{q} 
\nn \\
&+& (-)^{s(q+r)}\pbr{\ff{n-2}}{\ff{s}\we\ff{q}}\we\ff{r}
- \pbr{\ff{n-2}}{\ff{q}}\we\ff{r}\we\ff{s}
\\
&-&(-)^{q(r+s)}\pbr{\ff{n-2}}{\ff{r}}\we\ff{s}\we\ff{q}
- (-)^{s(q+r)}\pbr{\ff{n-2}}{\ff{s}}\we\ff{q}\we\ff{r} .
\nn
\eeqa
It is natural to call the algebraic structure appearing here a 
{\em higher-order Gerstenhaber algebra}. Similar to the 
Gerstenhaber algebra it has a graded commutative algebra structure 
-- that of the exterior algebra of horizontal forms, and the 
graded Lie algebra 
structure given by the bracket,  eqs. (7, 8). However, the graded derivation 
property of the bracket is generalized here as a higher-order Leibniz rule, 
eq. (10).

Further, it turns out that 
the implementation of  the  
Leibniz rule 
with respect to the exterior product implies a generalization of 
the map (5). This is related to the fact that the multivector field 
associated with a form according to eq. (5) exists only for a rather 
restricted class of forms which we call {\em Hamiltonian forms}. These forms 
depend in a specific polylinear way 
(the analysis of which will be presented elsewhere) on polymomenta 
so that their exterior product 
is not in general a Hamiltonian form. 
As the result, the higher-order  Leibniz rule with respect to the exterior 
product has no much sense if one is restricted to the subspace of Hamiltonian 
forms. 

Arbitrary horizontal forms 
are included by  mapping them 
to more general 
g.d.o.-s 
on the exterior algebra 
than those 
represented by vertical  multivectors in (5). 
In fact, any horizontal form $F$$\in \bigwedge^p_0({\cal Z})$ 
 can be mapped 
to a g.d.o. given by a vertical-multivector-valued 
form 
$\xt_F=\mu\otimes X$, where $\mu\in \bigwedge^1_0({\cal Z})$, 
$X\in V \! \bigwedge^{n-p+1}(T{\cal Z})$, 
so that
\beq
\xt_F \inn \Omega := \mu \we X\inn  \Omega =  \dv F.
\eeq
Note that the map (12), as well as (5), is degenerate 
 in the sense that 
there exist 
{\em primitive} g.d.o.-s $\xt_0$ which annihilate the polysymplectic form: 
$\xt_0\inn \Omega=0$, so that we, in fact, map horizontal forms to the 
equivalence 
classes  of g.d.o.-s on $\bigwedge^*({\cal Z})$ modulo an addition of 
primitive g.d.o.-s. 

Let us study the algebraic structure which appears now. 
First we define the 
bracket on g.d.o.-s to which dynamical variables (forms) are mapped and study 
its algebraic properties. The corresponding algebra will be a generalization 
of the Lie algebra of Hamiltonian vector fields in mechanics. 
We introduce the operation of ``lievization'' of a g.d.o. $X$ \cite{vinogr}: 
$L_X:=[X,\dv]=X\co \dv - (-)^{|X|}\dv \co X$, 
where $|X|$ is the degree of $X$, 
and define the {\em left semi-bracket} of g.d.o.-s  as follows: 
\beq
\lbr{X}{Y}:=[L_X,Y]=L_X\co Y - (-)^{|Y|(|X|+1)} Y\co L_X. 
\eeq
This operation has no obvious symmetry properties 
since the graded commutator of 
two g.d.o.-s is non-vanishing in general. Further, we call 
a g.d.o $X$ 
{\em locally Hamiltonian (LH)} if 
$L_X \Omega = 0$. Then, one can  straightforwardly prove that

{\bf Theorem 1.} {\em Locally Hamiltonian g.d.o.-s fulfill the axiom 
of the graded 
left Loday algebra}\footnote{We call graded  Loday algebra the 
graded analogue of   the Leibniz algebra  
(a non-commutative generalization of Lie algebras)  
introduced by Loday \cite{loday}. This choice 
seems to be reasonable to avoid a confusion with the Leibniz rule 
which can appear in closely related context (cf. [8a]).} 
{\em with respect to the left semi-bracket}:
\beq
\lbr{\lbr{X}{Y}}{Z}=\lbr{X}{\lbr{Y}{Z}}-
(-)^{(|X|+1)(|Y|+1)}\lbr{Y}{\lbr{X}{Z}}.
\eeq
Note that a weaker 
analogue of graded anticommutativity is valid for the 
semi-bracket as a consequence of (14):
$\lbr{\lbr{X}{Y}}{Z}=$$- (-)^{(|X|+1)(|Y|+1)} \lbr{\lbr{Y}{X}}{Z}$. 
In this sense the structure appearing here is a non-commutative analogue 
of a graded Lie algebra. 
 
The g.d.o.-s which satisfy eq. (12) are called {\em Hamiltonian g.d.o.-s.} 
They form a subalgebra 
in the graded Loday algebra of $LH$ g.d.o.-s. The semi-bracket of Hamiltonian 
g.d.o.-s induces a semi-bracket on horizontal forms:
$
-\dv \pbr{F}{G} = \lbr{X_F}{X_G}\inn \Omega ,
$
whence it follows
\beq
\pbr{F}{G}=(-)^{n-F}X_F \inn X_G \inn \Omega = (-)^{n-F}L_{X_F} G    .
\eeq
Again, by a straightforward calculation one proves

{\bf Theorem 2.} {\em The left semi-bracket equips the 
space of horizontal forms 
with the structure of the graded left Loday algebra , i.e. } 
\beq
\pbr{\pbr{F}{G}}{K}=
\pbr{F}{\pbr{G}{K}}- (-)^{(n-F-1)(n-G-1)}
\pbr{G}{\pbr{F}{K}}.
\eeq

Furthermore, the non-commutative graded Lie algebra 
structure above is  supplemented also by the graded analogue of 
the Poisson property:
 
{\bf Proposition.} {\em The following right graded Leibniz rule is fulfilled:}
\beq
\pbr{F\we G}{K}
=F\we \pbr{G}{K} + (-)^{G(n-K-1)}\pbr{F}{K}\we G. 
\eeq
For the proof one has to construct first the g.d.o. associated with $F\we G$: 
$X_{F\we G}=(-)^{G(F+1)}G \co X_F + (-)^F F \co X_G$. The 
rest is a straightforward calculation. 

In addition to the right graded Leibniz rule
the left higher-order graded 
Leibniz rule, eq. (10),  is also fulfilled. Thus, the algebraic 
structure on horizontal 
forms is a non-commutative analogue (in the sense of the non-commutativity 
of the  bracket operation) of the {\em right  Gerstenhaber algebra} 
(cf. e.g. \cite{flato}) and, 
at the same time,  the {\em left higher-order Gerstenhaber algebra}. 
This structure  generalizes the Poisson 
algebra structure of dynamical variables in mechanics to the 
DW Hamiltonian formulation of field theory. 
The graded Lie algebra of Hamiltonian forms 
treated earlier appears now as 
its
 (graded anti-commutative) sub-algebra.
Recall  that the graded Loday 
algebra structure was  found earlier in the context of the generalized 
BV algebras in \cite{ackman}.

In conclusion, note without proof that the bracket operation under 
discussion allows 
us to write the equations of motion of a dynamical variable 
$F\in \bigwedge^{*}_0({\cal Z})$ 
in Poisson bracket 
formulation [1]: 
$\bd F = \pbr{H\omega}{F}$, where 
$\bd F:=dx^i \we [\der_i y^a\der_a F 
+ \der_i p^j_a\der^a_j F]$
is the total vertical differential. 
This generalizes the well-known relationship between the total time 
derivative and the Poisson bracket with Hamilton's function in mechanics.
The DW equations, eq. (2), follow from  the  equation above if 
a form $F$ is chosen 
to be $p_a^i\om_i $ and 
$y^a$ respectively. 

Summarizing, we have discussed a possible intrinsic meaning of the 
polysymplectic form which was suggested earlier as a 
basic structure for the DW Hamiltonian formulation of field theory 
[2] (cf. [3b]). We have also considered an extension of the algebraic 
structure on the subspace of Hamiltonian forms in [2] to 
general horizontal forms. As the structures under consideration are 
field theoretic analogues of the structures  known to be essential for 
quantization procedures based on the 
structures of the 
standard 
Hamiltonian formalism, 
it would 
be interesting to extend the techniques of the 
canonical, deformation or geometric quantization to  the present 
framework. This could be both fascinating 
mathematical study and interesting physical problem  
a solution of which 
could contribute to our further 
understanding of quantum field theory and an interplay 
between relativity and quantum theory. 

{\bf Acknowledgments.} I thank 
Prof. H.D. Doebner and the Organizing Committee 
for the  invitation and 
for 
providing the support for attending the Colloquium. 
I gratefully acknowledge useful 
discussions with F. Cantrijn and Y. Kosmann-Schwarzbach 
at the Colloquium.


\end{document}